\documentclass[prd,aps,preprint,tightenlines,nofootinbib,superscriptaddress]{revtex4}
\pdfoutput=1
\usepackage{color}
\usepackage{amssymb,amsmath,psfrag,slashed,graphicx}
\usepackage{epsfig}
\usepackage{amsmath}
\usepackage{graphicx}
\usepackage{multirow}

\newcommand{\beq}{\begin{equation}}
\newcommand{\eeq}{\end{equation}}
\newcommand{\bqa}{\begin{eqnarray}}
\newcommand{\eqa}{\end{eqnarray}}

\begin{document}

\title{Graviton-mediated dark matter model explanation the DAMPE electron excess and search at $e^+e^-$ colliders}

\author{Ruilin Zhu$^{1}$~\footnote{rlzhu@njnu.edu.cn} and Yu Zhang$^{2,3}$~\footnote{Corresponding author:~dayu@nju.edu.cn} }

\affiliation{
$^1$ Department of Physics and Institute of Theoretical Physics,\\
Nanjing Normal University, Nanjing, Jiangsu 210023, China\\
$^{2}$ School of Physics,\\
Nanjing  University, Nanjing, Jiangsu 210039, China \\
$^3$ CAS Center for Excellence in Particle Physics, Beijing 100049, China}

\begin{abstract}
The very recent result of the DAMPE cosmic ray spectrum of electrons shows a narrow bump above
the background at around 1.4 TeV. We attempt to explain the DAMPE electron excess in a simplified Kaluza-Klein graviton-mediated
dark matter model, in which the graviton only interacts with leptons and dark matter.
The related phenomenological discussions are given and this  simplified graviton-mediated
dark matter model has the potential to be cross-tested in future lepton collider experiments.

\end{abstract}

\maketitle

{\it Introduction.} Very recently, the DArk Matter Particle Explorer (DAMPE~\cite{TheDAMPE:2017dtc}) Collaboration have reported
the first result of the  cosmic ray electrons flux up to 5 TeV, in which the electron excess over the background is shown around 1.4 TeV
~\cite{Ambrosi:2017wek}. This bump indicates the existence of the new sources to produce the electrons.

After the release of the DAMPE data, there are many interesting theoretical works
to explain the electron excess~\cite{Fan:2017sor,Fang:2017tvj,Duan:2017pkq,Gu:2017gle,Zu:2017dzm,Tang:2017lfb,Chao:2017yjg,
Athron:2017drj,Cao:2017ydw,Liu:2017rgs,Gu:2017bdw,Duan:2017qwj,Huang:2017egk,
Cholis:2017ccs,Chao:2017emq,Gao:2017pym,Niu:2017hqe}. These works to interpret of the DAMPE cosmic ray electron
excess can be assigned into two categories as follows.
One is to explain the excess from the continuous sources. For example, the dark matters annihilate into the standard model particles and then produce the DAMPE electron excess. The other one is to explain the excess from the burst-like source. For example, the pulsars can also be invoked to explain
  the origin of the high-energy electrons. Since the gravitational evidence for the dark matter is robust, it would be desirable if the DAMPE electron excess is related to the physics of the dark matter. In this paper, we will attempt to take  the dark matter annihilation as the source of the DAMPE electron excess.

Since the sharp peak is around 1.4TeV and is very close to  the TeV scale in the warped extra dimensions,
we will introduce the  Kaluza-Klein (KK) graviton excitation state, and take the KK graviton excitation state as
the mediator between the dark matter and standard model particles. The KK graviton can be reasonably
produced and are useful in many models such as the original Randall-Sundrum (RS) model ~\cite{Randall:1999ee},
bulk RS model~\cite{Agashe:2007zd,Fitzpatrick:2007qr}, and other warped extra dimension models~\cite{Han:1998sg}.
which can not only  produce the graviton at the TeV scale, but also provide  a solution to the hierarchy problem.
In this work, we consider simplified  graviton mediated dark matter model in which the KK graviton only interact with leptons and the  dark matter in order to
interpret of the DAMPE electron excess, and study the constraints on the model parameter space.

{\it Kaluza-Klein graviton excitation state.} The  KK graviton around TeV is introduced in many frameworks. In the brane-bulk scenario of the RS model,
the dimension of the spacetime is assumed to be D=4+1=5 with one compactified extra dimension. All the standard model particles were assumed to be localized on the IR-brane, while the gravity can propagate in
 the whole five-dimensional bulk. The bulk metric is
\begin{eqnarray}\label{matric}
  ds^2 &=& e^{-2{\cal K}R_c\phi}\eta_{\mu\nu}dx^\mu dx^\nu+R_c^2d\phi^2,
\end{eqnarray}
where $0\leq\phi\leq\pi$ and $\eta_{\mu\nu}$ denotes the flat Minkowski metric; $R_c$ denotes the compactified radius; and ${\cal K}$ is the curvature scale of the bulk.

After taking  a linear expansion of the gravity field as fluctuations around the flat metric and using
the  Kaluza-Klein reduction, many massive KK graviton excitation states are produced  in the effective four-dimensional theory. The masses of the KK graviton states can be expressed as
~\cite{Patrignani:2016xqp}
\begin{eqnarray}\label{mass}
  m^G_n &\simeq& (n+\frac{1}{4})\pi{\cal K}e^{-\pi{\cal K}R_c}  ,
\end{eqnarray}
where $n=1,2,\cdots$. From it, one can easily find $(m^G_2/m^G_1)^2\simeq 3.24$ and $(m^G_3/m^G_1)^2\simeq 6.76$.
Since the mass ratio between the higher excited state and the lightest excited state is large, the propagator suppression effect is obvious for low energy interaction processes.

 In the following, we will consider the simplified graviton-mediated dark matter model  where only one excited graviton $G^{KK}$ with the mass $m_G$ and decay width $\Gamma_G$ is introduced as many literatures did~\cite{Lee:2013bua,Lee:2014caa}. We attempt to explain the electron excess through the graviton as the mediator between the leptons and the dark matter, and consider the constraints to the parameters in the simplified graviton-mediated dark matter model.

{\it Graviton-mediated dark matter model.} The graviton-mediated dark matter model was proposed in
Refs.~\cite{Lee:2013bua,Lee:2014caa}, and is studied in many works~\cite{Han:2015cty,Kraml:2017atm,Dillon:2016tqp,Zhang:2016xtc}.
In this model, the dark sector do not directly couple to standard
model particles and is communicated with the standard
model particles through a spin-2 particle in warped extra-dimension. In this paper, we will consider the graviton-mediated dark matter model and restrict the graviton excitation state only coupling to the leptons in the standard model side. Besides, we will discuss the possibility of the fermion and vector types of dark matter.

The Graviton-mediated part of the Lagrangian  is
\begin{eqnarray}\label{La}
{\cal L}_{int}(x)&=& -\frac{1}{\Lambda}G^{KK}_{\mu\nu}(x)(c_D T^{\mu\nu}_D(x)+c_\ell T^{\mu\nu}_{\ell} (x)) \ ,
\end{eqnarray}
where $G^{KK}_{\mu\nu}(x)$ denotes the spin-2 KK graviton excitation field; $\Lambda$ is the interaction scale parameter; $c_D$ is the coupling parameter between the graviton and the dark matter and
$c_\ell$ is the coupling parameter between the graviton and the leptons in the standard model. Considering the experimental data which have indicated a new source couplings to the leptons,  we only
consider the graviton coupling to the leptons in the standard model particles side here. And for the dark side, we will consider the fermion and vector types dark matter. The energy-momentum tensor of the fermion dark matter is
\begin{eqnarray}\label{tensor}
T^{\mu\nu}_{D,\chi}&=& \frac{i}{4}\bar{\chi}(\gamma^\mu \partial^\nu+\gamma^\nu \partial^\mu)\chi-\frac{i}{4}( \partial^\nu\bar{\chi}\gamma^\mu+ \partial^\mu\bar{\chi}\gamma^\nu)\chi\nonumber\\&&-\eta^{\mu\nu}(i \bar{\chi}\gamma^\alpha \partial_\alpha\chi-m_\chi\bar{\chi}\chi)+\frac{i}{2}\eta^{\mu\nu}\partial_\alpha(\bar{\chi}\gamma^\alpha\chi) ,
\end{eqnarray}
The energy-momentum tensor of the vector dark matter is
\begin{eqnarray}\label{tensorv}
T^{\mu\nu}_{D,X}&=& \frac{1}{4}\eta^{\mu\nu}X^{\alpha\beta}X_{\alpha\beta}-X_{\alpha\lambda}X^\lambda_\beta+m_X^2(X_\mu X_\nu-\frac{1}{4}\eta^{\mu\nu}X^{\alpha}X_{\alpha} ),
\end{eqnarray}

Considering the spin-2 KK graviton, its propagator in the de Donder gauge can be expressed as
\begin{eqnarray}\label{pro}
\tilde{G}^{KK}_{\mu\nu\alpha\beta}&=& \frac{1}{2}D(s)[\eta_{\mu\alpha}\eta_{\nu\beta}+\eta_{\mu\beta}\eta_{\alpha\nu}
-\frac{2}{3}\eta_{\mu\nu}\eta_{\alpha\beta}] ,
\end{eqnarray}
 with
\begin{eqnarray}\label{pro2}
D(s)&=& \frac{i}{s-(m_G)^2+i{m_G} \Gamma_G} ,
\end{eqnarray}
where $\Gamma_G$ and $m_G$ are the total decay width and mass of the spin-2 KK graviton, respectively.

According to the Lagrangian in Eq.~(\ref{La}), the dark matter can couple to the leptons through the mediation of the spin-2 KK graviton. This mechanism may provide a new source to produce the leptons and then may explain the electron excess in the DAMPE experiment.
\begin{figure}[hbt]
\begin{center}
\includegraphics[width=0.5\textwidth]{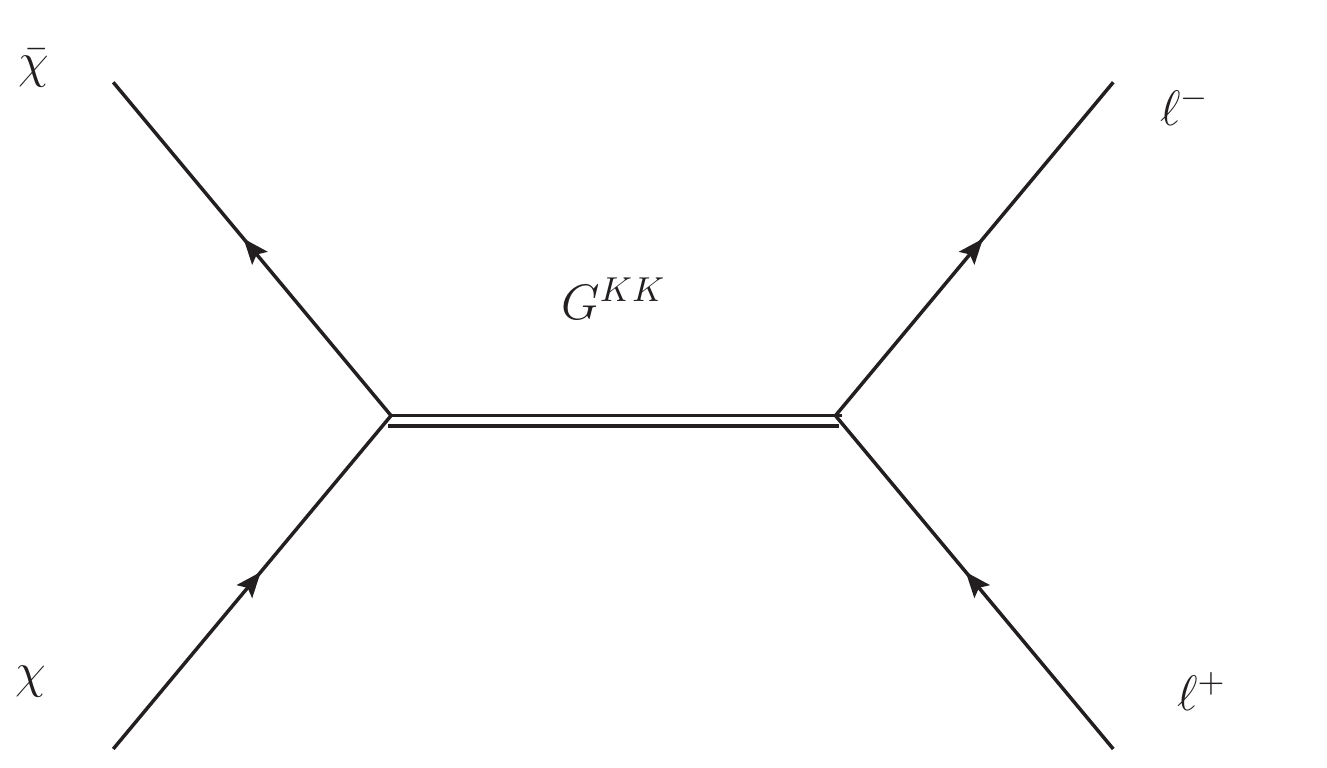}
\end{center}
\caption{Feynman diagrams for $\chi \bar{\chi}\to e^+ e^-$ where the graviton is the mediator between the dark mater and the standard model leptons. }
\label{fig:diagram}
\end{figure}

 The Feynman rules for the $G^{KK}_{\mu\nu}(k_3)-\bar{\psi}(k_1)-\psi(k_2)$ vertices among graviton and fermions are
 \begin{eqnarray}\label{feyrule}
-i\frac{c_\ell}{4\Lambda}[\gamma^\mu(k_1-k_2)^\nu+\gamma^\nu(k_1-k_2)^\mu-2\eta^{\mu\nu}(\slashed k_1-\slashed k_2-2m_\ell)] .
\end{eqnarray}
 The Feynman rules for the $G^{KK}_{\mu\nu}(k_3)-X^\rho(k_1)-X^\sigma(k_2)$ vertices among graviton and vector dark matter are
 \begin{eqnarray}\label{feyrule2}
-2i\frac{c_X}{\Lambda}[B^{\mu\nu\rho\sigma}m_X^2
+(C^{\mu\nu\rho\sigma\tau\beta}-C^{\mu\nu\rho\beta\sigma\tau})k_{1,\tau}k_{2,\beta}+\frac{1}{\xi}E^{\mu\nu\rho\sigma}(k_1,k_2)] ,
\end{eqnarray}
where
 \begin{eqnarray}\label{feyrule3}
B^{\mu\nu\alpha\beta}&=& \frac{1}{2}(\eta^{\mu\nu}\eta^{\alpha\beta}-\eta^{\mu\alpha}\eta^{\nu\beta}-\eta^{\mu\beta}\eta^{\nu\alpha}),\\
C^{\rho\sigma\mu\nu\alpha\beta}&=& \frac{1}{2}[\eta^{\rho\sigma}\eta^{\mu\nu}\eta^{\alpha\beta}-
(\eta^{\rho\mu}\eta^{\sigma\nu}\eta^{\alpha\beta}+\eta^{\rho\nu}\eta^{\mu\sigma}\eta^{\alpha\beta}+
\eta^{\rho\alpha}\eta^{\mu\nu}\eta^{\sigma\beta}+\eta^{\rho\beta}\eta^{\mu\nu}\eta^{\alpha\sigma})
],\\
E^{\mu\nu\rho\sigma}(k_1,k_2)&=&\eta^{\mu\nu}(k_1^\rho k_1^\sigma+k_2^\rho k_2^\sigma+k_1^\rho k_2^\sigma)-[\eta^{\nu\sigma}k_1^\rho k_1^\mu+\eta^{\nu\rho}k_2^\sigma k_2^\mu+(\mu\leftrightarrow \nu)].
\end{eqnarray}

The leading order Feynman diagram for $\chi \bar{\chi}\to e^+ e^-$ is plotted in Fig.~\ref{fig:diagram}. After computing the Feynman diagram in Fig.~\ref{fig:diagram}, one can obtain the cross section of the process $\chi (k_1) \bar{\chi}(k_2)\to G^{KK}\to e^+(p_1) e^-(p_2)$. For the fermion-type dark matter pair annihilation into electrons, we get the cross section
 \begin{eqnarray}\label{dsigma0}
\sigma(\chi \bar{\chi}\to e^+ e^- )&=& \frac{41c_\ell^2c_\chi^2s^{3/2}(-44m_\chi^4
+15m_\chi^2s+s^2 ) }{1152\pi\Lambda^4((s-m_G^2)^2+({m_G} \Gamma_G)^2)\sqrt{s-4m_\chi^2}} ,
\end{eqnarray}
where  we have ignored the small lepton mass.
The dark matter relative velocity is written as $\vec{v}=\frac{\vec{k_1}}{k^0_1}-\frac{\vec{k_2}}{k^0_2}$ and then $|\vec{v}|=2\sqrt{\frac{s-4m_\chi^2}{s}}$.
The fermion dark matter annihilation cross section into lepton pair multiplied by the dark matter relative velocity is
 \begin{eqnarray}\label{dsigma1}
\sigma|\vec{v}|(\chi \bar{\chi}\to e^+ e^- )&=& \frac{41c_\ell^2c_\chi^2 s(-44m_\chi^4
+15m_\chi^2s+s^2 )}{576\pi\Lambda^4((s-m_G^2)^2+({m_G} \Gamma_G)^2)} .
\end{eqnarray}

 The vector dark matter annihilation cross section into lepton pair multiplied by the dark matter relative velocity  can be computed similarly, which can be written as
 \begin{eqnarray}\label{dsigma2}
\sigma|\vec{v}|(X X \to e^+ e^-)&=& \frac{5c_\ell^2c_X^2 s^{2}(s-4m_X^2)}{864\pi\Lambda^4((s-m_G^2)^2+({m_G} \Gamma_G)^2)} \nonumber\\
&=& \frac{5c_\ell^2c_X^2 s^{3}|\vec{v}|^2}{3456\pi\Lambda^4((s-m_G^2)^2+({m_G} \Gamma_G)^2)}.
\end{eqnarray}
From the above expression, one can easily find that the cross-section  multiplied by the vector dark matter relative velocity has a suppression factor $|\vec{v}|^2$ and then will become trivial when $|\vec{v}|\to 0$. Thus the fermion dark matter is more better choice to explain the DAMPE electron excess.

So we will adopt the fermion type dark matter to explain the DAMPE electron excess. In the graviton-mediated dark matter model, there are only four independent parameters, two masses and two couplings:
\beq
{m_\chi, m_G,c_\ell/\Lambda, c_\chi/\Lambda}.
\eeq

The total  decay width of the spin-2 graviton into the leptons and dark matter reads
\beq
\Gamma_G = \sum_{\ell=e,\mu,\tau,\nu_e,\nu_\mu,\nu_\tau}\Gamma_G |_{\ell\bar\ell}+\Theta (m_G-2m_\chi)\Gamma_G |_{\chi\bar\chi},
\eeq
 where
  \begin{eqnarray}\label{gamma}
 \Gamma_G|_{\ell\bar\ell}&=& \frac{m_G^3c_\ell^2}{80\pi\Lambda^2},
 \end{eqnarray}
 and
  \begin{eqnarray}\label{gamma}
 \Gamma_G|_{\chi\bar\chi}&=& \frac{m_G^3c_\chi^2}{80\pi\Lambda^2}(1-4\frac{m_\chi^2}{m_G^2})^{3/2}(1+\frac{8m_\chi^2}{3m_G^2}),
 \end{eqnarray}
 which has been calculated in Ref.~\cite{Han:1998sg}.

\begin{figure}[hbt]
\begin{center}
\includegraphics[width=0.6\textwidth]{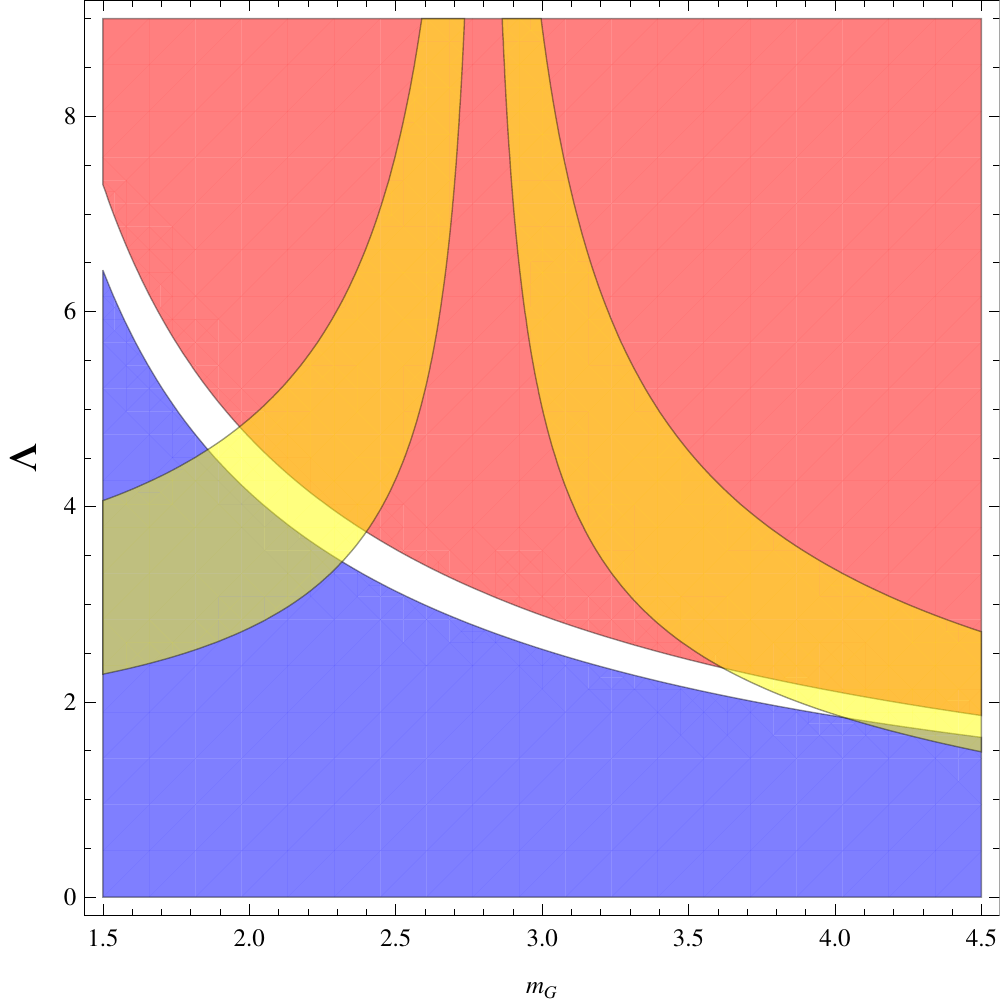}
\end{center}
\caption{
 The yellow region stands for the DM annihilation cross section multiplied by the relative velocity in $[3\times10^{-26},3\times10^{-25}]$ cm$^3$s$^{-1}$. The red and blue regions stand for significance less than $ 3$ and  larger than $5$ of  the graviton-mediated model search at 1TeV ILC with 1000 $fb^{-1}$ luminosity through $e^+e^-\to G^{KK}\to \mu^+\mu^-$ channel, respectively.
   }
\label{fig:fit}
\end{figure}

In this work, we set $c_\ell=c_\chi=1$ and $m_\chi=1.4$ TeV and can
get the DM annihilation cross section as the function of $\Lambda$ and $m_G$.
In Fig. \ref{fig:fit}, the yellow region stands for the  DM annihilation cross section  multiplied by the relative velocity
in $[3\times10^{-26},3\times10^{-25}]$ cm$^3$s$^{-1}$ which can interpret of the DAMPE electron-positron excess \cite{Fan:2017sor}. From it, one can get a large parameter space in the graviton-mediated dark matter model to explain the excess successfully. In the following, we will consider the constraint of  the parameters at $e^+e^-$ colliders.

\begin{figure}[hbt]
\begin{center}
\includegraphics[width=0.8\textwidth]{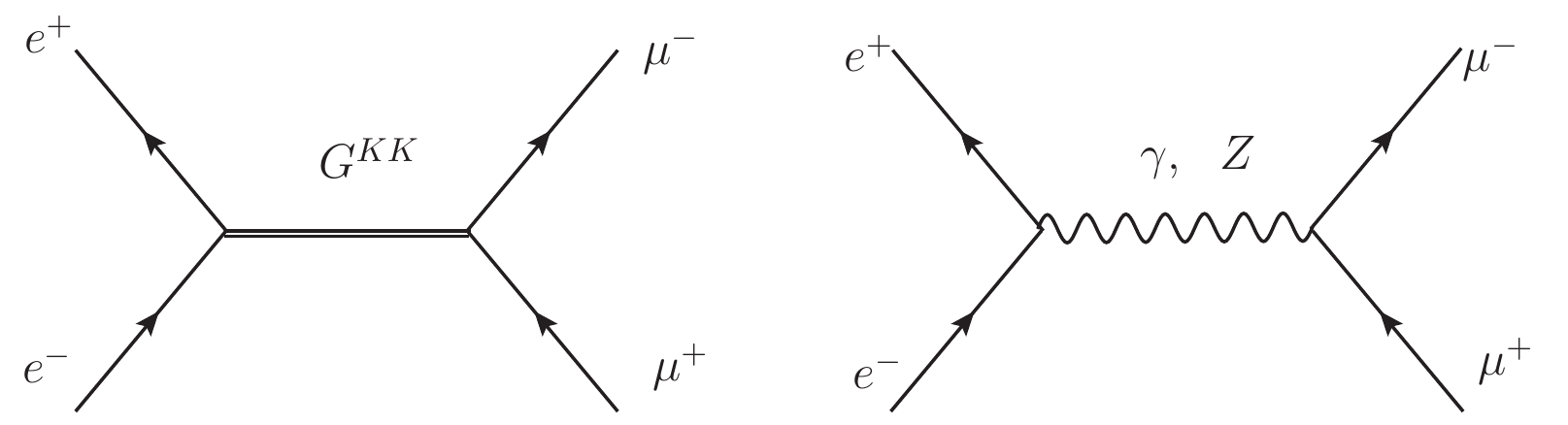}
\end{center}
\caption{Feynman diagrams for $e^+ e^-\to \mu^+ \mu^-$ with the mediator from the graviton and the standard model gauge bosons. }
\label{fig:eemm}
\end{figure}

{\it Search for the spin-2 KK graviton at $e^+e^-$ colliders.}
Since the graviton only interacts with leptons, we may  get constraints from
measurements on the cross section of $e^+e^-\to \ell^+\ell^-$.
In this section, we will look at the possible effects at  future $e^+e^-$ colliders, such as CEPC
 \cite{CEPC-SPPCStudyGroup:2015csa}, ILC \cite{Baer:2013cma}
and FCC-ee \cite{Gomez-Ceballos:2013zzn}.
For the process  $e^+e^-\to e^+e^-$, the enhancement t-channel will make the standard model predictions
large. The possible new physics effect from the graviton is hard to detect  in the large background.
For the process  $e^+e^-\to \mu^+\mu^-$, besides the $Z/\gamma$-mediated diagrams (at the right of Fig. \ref{fig:eemm})
, there will introduce additional graviton-mediated diagram (at the right of Fig. \ref{fig:eemm}).
Therefore, the cross section in the graviton-mediated model can be written as
  \begin{eqnarray}
\sigma|_{e^+e^-\to \mu^+\mu^-} &=& \left| (M|_{e^+e^-\to Z/\gamma\to \mu^+\mu^-} +
M|_{e^+e^-\to G\to \mu^+\mu^-} )\right|^2
\nonumber\\&\simeq & \sigma |_{e^+e^-\to Z/\gamma\to \mu^+\mu^-}+
\sigma |_{e^+e^-\to G\to \mu^+\mu^-},
  \end{eqnarray}
where we have found that the interference between the $Z/\gamma$-mediated and graviton-mediated is trivial.

The prediction of the cross section in the standard model is
 \begin{eqnarray}\label{gamma}
\sigma|_{e^+e^-\to Z/\gamma\to \mu^+\mu^-}&=& \pi  \alpha ^2[\frac{ 8 C_w^4 \left(17 s m_Z^2+8 m_Z^4+3 s^2\right)-64 C_w^6 m_Z^2 \left(2 m_Z^2+s\right)}{48 s C_w^4 \left(C_w^2-1\right){}^2 \left(s-m_Z^2\right){}^2}\nonumber\\
   &&\frac{ -24 s C_w^2 \left(3 m_Z^2+2 s\right)+64 C_w^8 m_Z^4+25
   s^2}{48 s C_w^4 \left(C_w^2-1\right){}^2 \left(s-m_Z^2\right){}^2}].
\end{eqnarray}
This gives the cross section of the channel $e^+e^-\to \mu^+\mu^-$ to be around (2.0,1.8,0.93,0.45,0.11) pb
at the center-of-mass energy $\sqrt{s}=(240,250,350,500,1000)$GeV, respectively.

The prediction of the cross section from the graviton is
 \begin{eqnarray}\label{gamma}
\sigma|_{e^+e^-\to G\to \mu^+\mu^-}&=& \frac{41c_l^4 s^3}{1152\pi\Lambda^4((s-(m^G)^2)^2+({m^G} \Gamma^G)^2)}.
\end{eqnarray}

To further explore the research potential for the graviton-mediated  signals at the $e^+e^-$ colliders,
we define the significance (S) as
\beq
 S=\frac{\sigma|_{e^+e^-\to G\to \mu^+\mu^-} \times L }
{\sqrt{\sigma|_{e^+e^-\to Z/\gamma\to \mu^+\mu^-}\times L}},
\eeq
 where $L$ is the integrated luminosity
of the collider. We calculate the significance at the CEPC with $\sqrt s=240$, at FCC-ee with $\sqrt s=350$ GeV and
at ILC with $\sqrt s=1000$ GeV. The integrated luminosities are all assumed to be 1000 fb$^{-1}$.
We find that the significance can not be more than 3 in the plotted region of Fig. \ref{fig:fit} at CEPC and FCC-ee.
In the Fig. \ref{fig:fit}, we present the result of significance obtained at 1 TeV ILC with 1000 fb$^{-1}$ luminosity,
$S\ge 5$ region is plotted in blue and $S\le 3$ region is plotted in red.
We can see that some ``DAMPE favored" region may be discoverable in the future 1 TeV ILC.
We select a benchmark model point as follows ($m_\chi$, $m_G$, $c_l/\Lambda$, $c_\chi/\Lambda$)
= (1.4 TeV, 1.8 TeV, 1/4.2 TeV$^{-1}$, 1/4.2 TeV$^{-1}$). The DM annihilation for this benchmark model
is 1.3 pb, which can explain the DAMPE excess and relic density.
The $\sigma|_{e^+e^-\to G\to \mu^+\mu^-}$ is 2.83 fb, which may be able to searched in the future.

{\it Conclusions.}
In this paper, we propose a simplified KK graviton-mediated dark matter model
in which the graviton only couples with leptons and the dark matter and explain the
electron plus positron excess at energies around 1.4 TeV recently observed by the DAMPE experiment.
We introduce the fermion and vector types dark matter respectively, and we find that the vector dark matter annihilation cross section into lepton pair multiplied by the dark matter
relative velocity will vanish when $|\vec{v}|\to 0$ and the Dirac fermion dark matter can accommodate the
DAMPE excess by setting the suitable parameters.
We also present the related phenomenological discussions at future $e^+e^-$ colliders,
and find that this simplified graviton-mediated dark matter model  may be discoverable at the ILC with $\sqrt s=1000$ GeV
and 1000 fb$^{-1}$ luminosity.

{\it Acknowledgments.}
This work was supported in part by the National Natural Science Foundation
of China under Grant No. 11647163 and 11705092, by Natural Science Foundation of
Jiangsu under Grant No. BK20171471 and by the China Postdoctoral Science Foundation under Grant No.2017M611771.

\end{document}